\documentclass[aps,prd,twocolumn,preprintnumbers,superscriptaddress,nofootinbib]{revtex4}
\pdfoutput=1

\newcommand{\PRE}[1]{}				

\usepackage{amsmath}
\usepackage{amsfonts}
\usepackage{amssymb}
\usepackage{epsfig}
\usepackage{graphicx}
\usepackage{setspace}

\newcommand{\beq}{\begin{equation}}
\newcommand{\eeq}{\end{equation}}

\newcommand{\wjmL}{\left(
                         \begin{array}{ccc}
       l_1 & l_2  & L  \\
         m_1 & m_2  & M
                         \end{array}
                   \right)}
\newcommand{\wjmmL}{\left(
                         \begin{array}{ccc}
       l_3 & l_4  & L  \\
         m_3 & m_4  & -M
                         \end{array}
                   \right)}

\newcommand{\wj}{\left(
                         \begin{array}{ccc}
       l_1 & l_2  & l_3  \\
         0 & 0  & 0
                         \end{array}
                   \right)}

\begin{document}
\title{\PRE{\vspace*{1.5in}}
Constraints on Spatial Variations in the Fine-Structure constant from Planck
\PRE{\vspace*{0.3in}}}

\author{Jon O'Bryan\footnote{ jobryan@uci.edu}}
\affiliation{Center for Cosmology, Department of Physics and Astronomy,
University of California, Irvine, CA 92697, USA }

\author{Joseph Smidt\footnote{ jsmidt@uci.edu}}
\affiliation{Center for Cosmology, Department of Physics and Astronomy,
University of California, Irvine, CA 92697, USA }

\author{Francesco De Bernardis\footnote{ fdeberna@uci.edu}}
\affiliation{Center for Cosmology, Department of Physics and Astronomy,
University of California, Irvine, CA 92697, USA }
%
%
\author{Asantha Cooray\footnote{acooray@uci.edu}} 
\affiliation{Center for Cosmology, Department of Physics and Astronomy,
University of California, Irvine, CA 92697, USA }
%


\date{\today}

\begin{abstract}
\PRE{\vspace*{.3in}}
We use the Cosmic Microwave Background (CMB) anisotropy data from Planck  to constrain the spatial fluctuations of the fine-structure constant $\alpha$. 
Through Thompson scattering of CMB photons, spatial anisotropies of $\alpha$ lead to higher-order correlations in the CMB anisotropies.
We use a quadratic estimator based on the four-point correlation function of the CMB temperature anisotropy to extract the angular power spectrum of the 
spatial variation of the fine-structure constant projected along the line of sight at the last scattering surface. At tens of degree angular scales and above,
we constrain the rms fluctuations of the fine structure constant to be $\delta \alpha/\alpha_0= (1.34 \pm 5.82) \times 10^{-2}$ at the $95\%$ confidence level with respect to the standard 
value $\alpha_0$. We find no evidence for a spatially varying $\alpha$ at a redshift of $10^3$.
\end{abstract}


\maketitle


One of the key questions of modern  physics concerns the possibility that physical constants vary across space and time in the history of the universe. 
One possible variation that has received recent attention is that of the fine structure constant, $\alpha$. The standard value of $\alpha$  from 
measurements of the electron magnetic moment anomaly is $\alpha_0 = 1/137.035999074(44)$ \cite{CODATA}, however there have been reports of statistically significant variations in this constant from 
high redshifts in quasar absorption line systems ($\delta \alpha/\alpha_0 = (-0.72\pm 0.18) \times 10^{-5}$ ~\cite{Webb98, Webb01}), though
others find such variations to be insignificant ($\delta \alpha/ \alpha_0 = (-0.6 \pm 0.6) \times 10^{-6}$ \cite{Murphy:2003, Srianand:2004}).

The Cosmic Microwave Background (CMB) temperature anisotropies are now known to be
a sensitive probe of the fine structure constant  given Thompson scattering of CMB photons \cite{Nakashima:2008}.  The CMB data has thus been extensively used already to constrain the mean
value of $\alpha$ at the last scattering surface and to search for any time variation of $\alpha$ between a redshift of 10$^3$ and today~\cite{Rocha:2003, Martins:2004, Ichikawa:2006, Stefanescu:2007, Nakashima:2008,Menegoni:2009rg, Menegoni:2012tq}. Such searches are motivated by the theoretical models that predict the possibility that fine-structure constant may vary with time. Such models range from higher 
dimensional Kaluza-Klein theories (see \cite{Uzan:2003} for a review) to long-range forces coupling a scalar axion to photons \cite{Bekenstein:2002}, e-Brane cosmology \cite{Youm:2002}, and varying-$e$ (electron charge) models \cite{Sandvik:2002}. A time-variation of $\alpha$, naturally, also implies spatial variations or anisotropies of $\alpha$ from one region of the universe to another. 
Independent studies of quasar absorption lines using the Keck telescope \cite{Murphy:2004} and the UVES (Ultraviolet and Visual Echelle Spectrograph) on the VLT (Very Large Telescope) have detected a non-zero
spatial variation of $\alpha$ in the form of a dipole with a statistical significance of 4.2$\sigma$ \cite{King:2012}. If this is the case, theoretical models also suggest that spatial fluctuations
must exist in higher orders of $\alpha$ and not just in the form of a dipole.

Here we use the trispectrum \cite{Hu:2001} of the recent Planck CMB data \cite{Planck13} to constrain the spatial dependence of the fine structure constant at the last scattering surface. 
Following the early works of Ref.~\cite{Sigurdson:2009jr} that predicted spatial fluctuations of $\alpha$,
we make use of a revised all-sky estimator based on the four-point correlation function of CMB maps. 

\begin{figure}
    \begin{center}
      {\includegraphics[scale=0.45]{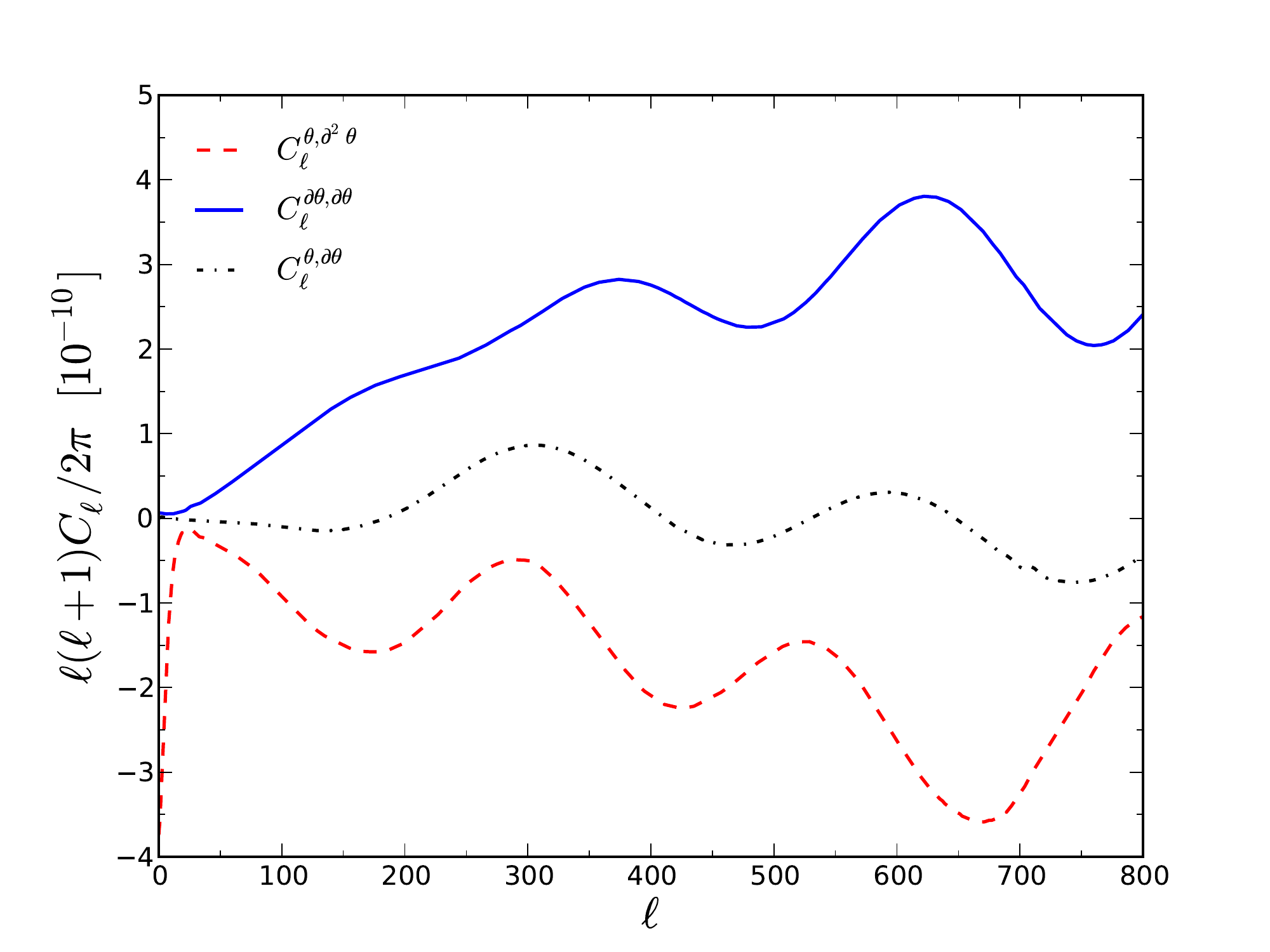} 
       }
   \end{center}
   \vspace{-0.7cm}
   \caption[width=3in]{Plot of $C_{\ell}^{\partial\theta\partial\theta}$ (solid; assuming $\delta \alpha/\alpha = 0.08$),  $C_{\ell}^{\partial\theta\partial\theta}$ (dashed dotted; assuming $\delta \alpha/\alpha = 0.01$), and $C_{\ell}^{\theta\partial^2\theta}$ (dashed; assuming $\delta \alpha/\alpha = 0.01$) derivative power spectra for Planck best fit parameters.}   
   \label{fig:derivps1}
\end{figure}


In the standard cosmological scenario, the frequency of oscillations in the primordial photon-baryon plasma imprinted on the CMB power spectrum depend on a visibility function that describes the probability density for a photon to last scatter at redshift $z$. This visibility function is a function of the fine-structure constant since $\alpha$ determines the fraction of free electrons as a function of time.  Therefore, $\alpha$ is a key parameter of the the ionization history through the Thomson scattering processes. It is clear that a variation of $\alpha$ affects the recombination by changing the shape and shifting in time the visibility function, which in turn affect the shape and position of the peaks of the CMB angular power spectrum.

To calculate the observable effects of a spatially dependent $\alpha$ on the CMB temperature map we follow an approach similar to Ref.~\cite{Sigurdson:2009jr}. We first perform a spherical harmonics expansion of the temperature field $\theta$:
\begin{eqnarray}\label{eq:theta}
\tilde{\theta}_{\ell m} &\approx& \theta_{\ell m} + \displaystyle\int \, dn \, Y_{\ell m}^* \delta\alpha \frac{\partial \theta}{\partial \alpha} \\ \nonumber
		& & \quad + \frac{1}{2} \displaystyle\int \, dn \, Y_{\ell m}^* (\delta \alpha)^2 \frac{\partial^2 \theta}{\partial \alpha^2} \\
		& = & \theta_{\ell m} + \displaystyle\sum_{\ell_1 m_1, \ell_2 m_2} \delta \alpha_{\ell_1 m_1} \left[  \left(\frac{\partial \theta}{\partial \alpha}\right)_{\ell_2 m_2}  I_{\ell \ell_1 \ell_2}^{m m_1 m_2} \right. \\ \nonumber
		& & \quad \left. + \frac{1}{2} \left(\frac{\partial^2 \theta}{\partial \alpha^2}\right)_{\ell_2 m_2} \displaystyle\sum_{\ell_3 m_3} \delta\alpha_{\ell_3 m_3}^* J_{\ell \ell_1 \ell_2 \ell_3}^{m m_1 m_2 m_3} \right]
\end{eqnarray}
where the $Y_{\ell m}$ are the spherical harmonics functions and the two integrals $I$ and $J$ are given by
\begin{eqnarray}
I^{m m_1 m_2}_{\ell \ell_1 \ell_2} &=& \displaystyle\int \, dn \, Y_{\ell m}^* Y_{\ell_1 m_1}^* Y_{\ell_2 m_2}^* \\
J^{m m_1 m_2 m_3}_{\ell \ell_1 \ell_2 \ell_3} &=& \displaystyle\int \, dn \, Y_{\ell m}^* Y_{\ell_1 m_1}^* Y_{\ell_2 m_2}^* Y_{\ell_3 m_3}^* \, ,
\end{eqnarray}\label{IJ}
respectively.

It can be shown that, retaining first-order corrections, no variations are present in the two-point (power spectrum) or three-point (bispectrum) correlation functions. We thus focus on the effects on the four-point correlation function (trispectrum).
%
%
%
%
%
\begin{figure}[t]
    \begin{center}
      { \includegraphics[scale=0.45]{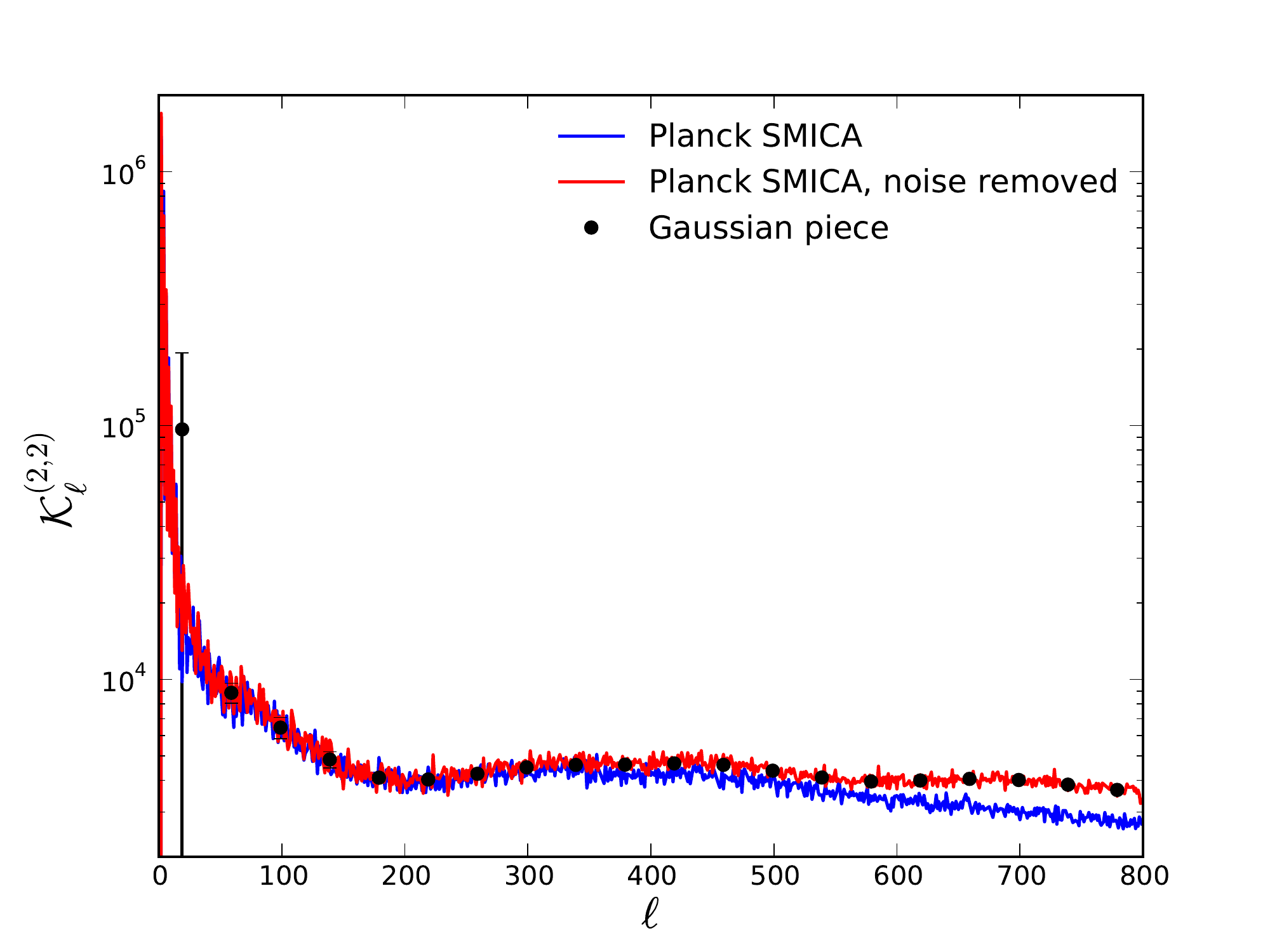} }
   \end{center}
   \vspace{-0.7cm}
   \caption[width=3in]{The estimator $K_\ell^{(2,2)}$ for Planck SMICA full sky data (blue) and Planck full sky data with noise removed SMICA map 
(red) compared to that obtained from full sky Gaussian simulations (black).}
   \label{fig:k22vg22}
\end{figure}
%

%

%
The Fourier counterpart of the four-point correlation function can be written as the sum of the Gaussian component and the \textit{connected} term as follows
\begin{eqnarray}\label{eq:tripieces}
\left<a_{l_1m_1} a_{l_2m_2} a_{l_3m_3} a_{l_4m_4}\right> &=& \\
\left<a_{l_1m_1} a_{l_2m_2} a_{l_3m_3} a_{l_4m_4}\right>_G\,   
&+& \left<a_{l_1m_1} a_{l_2m_2} a_{l_3m_3} a_{l_4m_4}\right>_c \, , \nonumber
\end{eqnarray}
where the $a_{\ell m}$ are the coefficients of the spherical harmonic expansion. The connected term of the Fourier transform, that is, the term remaining after the Gaussian component is subtracted in Eq.~\ref{eq:tripieces}, represent the trispectrum. The Gaussian and connected pieces can be expanded as
\begin{eqnarray}\label{eq:triconnected}
\left <a_{l_1m_1} a_{l_2m_2} a_{l_3m_3} a_{l_4m_4}\right>_G =&& \\ 
\sum_{L M} (-1)^{M} G_{l_1 l_2}^{l_3 l_4} (L)  \wjmL &&  \hspace{-0.5cm}  \wjmmL, \nonumber \\
\left <a_{l_1m_1} a_{l_2m_2} a_{l_3m_3} a_{l_4m_4}\right>_c =&& \\ 
\sum_{L M} (-1)^{M} T_{l_1 l_2}^{l_3 l_4} (L)  \wjmL &&  \hspace{-0.5cm}  \wjmmL, \, , \nonumber 
\end{eqnarray}
where the quantities in parentheses are the Wigner-3j symbols. The two functions $G_{l_1 l_2}^{l_3 l_4} (L)$ and $T_{l_1 l_2}^{l_3 l_4} (L) $ 
 for the Gaussian and connected components, respectively, can be calculated analytically. 
Proceeding from the expansion Eq.~\ref{eq:theta}, after some tedious but straightforward algebra, we arrive at
\begin{align}
	G_{\ell_3 \ell_4}^{\ell_1 \ell_2} (L) &= (-1)^{\ell_1 + \ell_3} \sqrt{ (2\ell_1 + 1)(2\ell_3 +1)} \\ \nonumber & \quad \times C_{\ell_1} C_{\ell_3} \delta_{L0} \delta_{\ell_1 \ell_2} \delta_{\ell_2 \ell_3} \\ \nonumber & \quad + (2L+1) C_{\ell_1} C_{\ell_2} \\ \nonumber & \quad \times \left[ (-1)^{\ell_2 + \ell_3 + L} \delta_{\ell_1 \ell_3} \delta_{\ell_2 \ell_4} + \delta_{\ell_1 \ell_4} \delta_{\ell_2 \ell_4} \right] \, ,
\end{align}
and
\begin{align}
T^{\ell_1\ell_2}_{\ell_3\ell_4,conn} &= C^{\alpha\alpha}_L F_{\ell_2 L \ell_1} F_{\ell_4 L \ell_3} \\ \nonumber
		&  \quad \times (C^{\theta \partial\theta/\partial\alpha}_{\ell_1}  + C^{\theta \partial\theta/\partial\alpha}_{\ell_2} ) (C^{\theta \partial\theta/\partial\alpha}_{\ell_3} + C^{\theta \partial\theta/\partial\alpha}_{\ell_4} ) \, ,
\end{align}
where
\begin{align}
F_{\ell_1 \ell_2 \ell_3} = \sqrt{\frac{(2\ell_1+1)(2\ell_2+1)(2\ell_2+3)}{4\pi}} \wj.
\end{align}
In these calculations, we used a modified version of \texttt{camb}  \cite{camb} to calculate the angular cross-correlation functions $C_{\ell}^{\theta,\partial\theta}$. Figure~\ref{fig:derivps1} shows the derivative power spectra $C_{\ell}^{\theta\partial^2\theta}$, $C_{\ell}^{\partial\theta,\partial\theta}$, and $C_{\ell}^{\partial\theta\partial\theta}$.
%
%
%

\begin{figure}[t]
    \begin{center}
      { \includegraphics[scale=0.45]{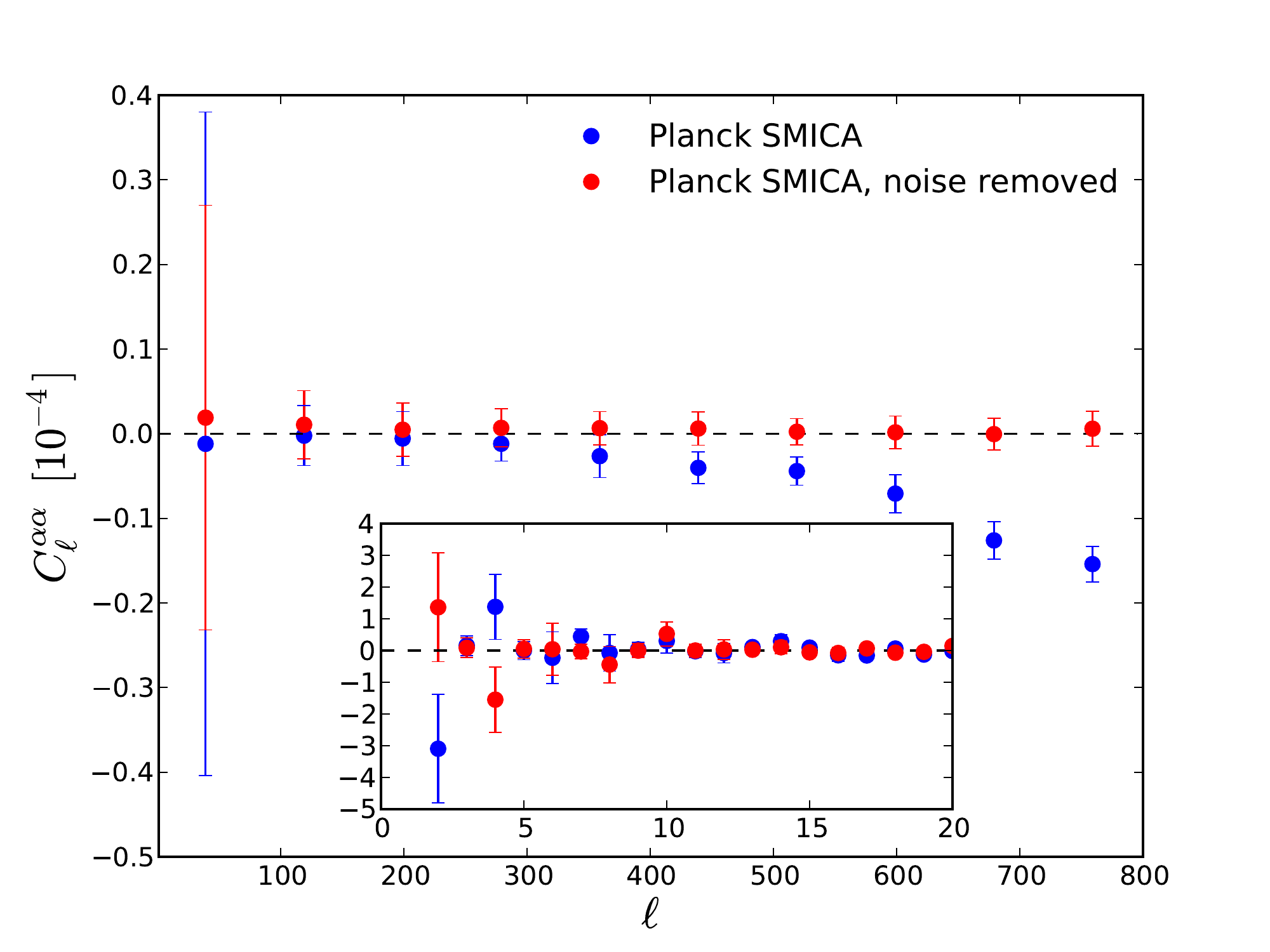} }
   \end{center}
   \vspace{-0.7cm}
   \caption[width=3in]{Power spectrum of spatial anisotropies of projected fine structure constant fluctuations at the last scattering surface.
The value of $C_{\ell}^{\alpha\alpha}$ is consistent with zero at $2\sigma$ for $\ell<5$ and at $1\sigma$ for the higher multipoles for noise removed maps. The inset shows
 the low-multipoles range without binning to highlight the fluctuations. We show two sets of measurements here using the Planck SMICA map  (blue) and the noise-removed SMICA map (red).
The detections at $\ell > 600$ is a result of the noise bias and is removed when using the noise-removed SMICA map. We find no statistically significant detection of $\alpha$ spatial
anisotropies once accounting for noise and other instrumental effects in Planck data.}
   \label{fig:claa}
\end{figure}

For simplicity we rewrite the trispectrum using the ansatz
\begin{align}
	T^{(i)\ell_1 \ell_2}_{\ell_3\ell_4}(L) &= h_{\ell_2 L \ell_1} h_{\ell_4 L \ell_3} F_L^{(i)} \alpha_{\ell_1}^{(i)} \beta_{\ell_2}^{(i)} \gamma_{\ell_3}^{(i)} \delta_{\ell_4}^{(i)}
\end{align}
where the functions  $\alpha_{\ell},    \beta_{\ell}, \gamma_{\ell}, \delta_{\ell}$ are given in Table~\ref{tab:weightings}.
\begin{table}[htbp]  \centering  
	\begin{tabular}{ | c | c | c | c | c |}
		\hline
		i & 1 & 2 & 3 & 4 \\ \hline
		$F_L$ & $C^{\alpha\alpha}_L$ & $C^{\alpha\alpha}_L$ & $C^{\alpha\alpha}_L$ & $C^{\alpha\alpha}_L$  \\ \hline
		$\alpha_{\ell_1}$ & $C^{\theta,\partial\theta/\partial\alpha}_{\ell_1}$ & $C^{\theta,\partial\theta/\partial\alpha}_{\ell_1}$ & $1$ & $1 $ \\ \hline
		$\beta_{\ell_2}$ & $1$ & $1$ & $C^{\theta,\partial\theta/\partial\alpha}_{\ell_2}$ & $C^{\theta,\partial\theta/\partial\alpha}_{\ell_2}$ \\ \hline
		$\gamma_{\ell_3}$ & $C^{\theta,\partial\theta/\partial\alpha}_{\ell_3}$ & $1$ & $C^{\theta,\partial\theta/\partial\alpha}_{\ell_3}$ & $1 $ \\ \hline
		$\delta_{\ell_4}$ & $1$ & $C^{\theta,\partial\theta/\partial\alpha}_{\ell_4}$ & $1$ & $C^{\theta,\partial\theta/\partial\alpha}_{\ell_4}$ \\ 
		\hline
    \end{tabular}  
    \caption{Weightings for trispectrum estimator.}  
\label{tab:weightings}
\end{table}
Analogously to Ref.~\cite{Smidt:2011}, the analytical and data trispectra estimator can be written as
\begin{align}
	\label{eq:k22ana}
	\mathcal{K}_{\ell,ana}^{(2,2)} &= \frac{1}{(2\ell + 1)} \displaystyle\sum_{\ell_i} \frac{1}{(2\ell + 1)} \frac{T_{\ell_1 \ell_2}^{\ell_3 \ell_4}(\ell) \hat{T}_{\ell_1 \ell_2}^{\ell_3 \ell_4}(\ell)}{\mathcal{C}_{\ell_1}\mathcal{C}_{\ell_2}\mathcal{C}_{\ell_3}\mathcal{C}_{\ell_4}} \\
	\label{eq:k22data}
 	\mathcal{K}_{\ell,data}^{(2,2)} &= \frac{1}{(2\ell +1)} \displaystyle\sum_m \left[ A^{(x)} B^{(x)} \right]_{\ell m} \left[ G^{(x)} D^{(x)} \right]_{\ell m}  \, .
\end{align}
\indent
The functions in square parenthesis in (\ref{eq:k22data}) are:
\begin{align}
	A^{(x)}_{\ell m} \equiv \frac{\alpha_{\ell}^{(x)}}{\mathcal{\tilde{C}}^{\ell}} b_{\ell} a_{\ell m}, \quad
	B^{(x)}_{\ell m} \equiv \frac{\beta_{\ell}^{(x)}}{\mathcal{\tilde{C}}^{\ell}} b_{\ell} a_{\ell m}, \\ 
	G^{(x)}_{\ell m} \equiv \frac{\delta_{\ell}^{(x)}}{\mathcal{\tilde{C}}^{\ell}} b_{\ell} a_{\ell m}, \quad
	D^{(x)}_{\ell m} \equiv \frac{\gamma_{\ell}^{(x)}}{\mathcal{\tilde{C}}^{\ell}} b_{\ell} a_{\ell m} \, ,
\end{align}
where $b_{\ell}$ is the beam transfer function, $a_{\ell m}$ are the Fourier coefficients for the data, $\tilde{C}^{\ell}$ contains noise, beam, and masking effects, and $\alpha_{\ell},    \beta_{\ell}, \gamma_{\ell}, \delta_{\ell}$ are listed in Table \ref{tab:weightings}. We calculate $\tilde{C}^{\ell}$ with a modified version of \texttt{camb}. 

The analytical forms of the Gaussian and connected estimators are respectively
\begin{align}
\label{eq:g22}
\mathcal{G}_{\ell}^{(2,2)} &=  \frac{1}{2 \pi} \displaystyle\sum_{\substack{\ell_1\ell_2 \\ \mod(\ell_1 + \ell_2 + L,2) = 0}} \frac{ C_{\ell}^{\alpha\alpha}  (C_{\ell_1}^{\theta\partial\theta} + C_{\ell_2}^{\theta\partial\theta})^2}{\tilde{C}_{\ell_1} \tilde{C}_{\ell_2}} \\ \nonumber
 & \quad \times (2\ell_{1}+1) (2\ell_{2}+1) \begin{pmatrix} \ell_1 & L & \ell_2 \\ 0 & 0 & 0 \end{pmatrix}^2 \, , \\ \nonumber
\end{align}
and
\begin{align}
\label{eq:k22conn}
	\mathcal{K}^{(2,2)}_{\ell,conn} 
&=\frac{1}{(2\ell+1)^2} \displaystyle\sum_{\ell_i m_i, L M} \frac{ C_{\ell}^{\alpha\alpha} F_{\ell_2 L \ell_1}^2 F_{\ell_4 L \ell_3}^2}{\tilde{C}_{\ell_1}\tilde{C}_{\ell_2}\tilde{C}_{\ell_3}\tilde{C}_{\ell_4}} \\ \nonumber
& \quad \times (C_{\ell_1}^{\theta\partial\theta}  + C_{\ell_2}^{\theta\partial\theta})^2 (C_{\ell_3}^{\theta\partial\theta} + C_{\ell_4}^{\theta\partial\theta} )^2 \, ,
\end{align}
where Eq.~\ref{eq:k22conn} is obtained by subtracting Eq.~\ref{eq:g22} from Eq.~\ref{eq:k22data}.
%
%
%
%
The estimators shown in the previous sections must be corrected for the effects of instrumental noise, beam and masking. In general the effect of the beam and the instrumental noise is to reduce the signal-to-noise ratio for the trispectrum estimator, while the masking introduces cut-sky mode-coupling effects that must be corrected for from the final power spectrum of $\alpha$.
To account for realistic observational and instrumental effects we need to add a beam modified factor to our power spectrum as follows
\beq
C_\ell\rightarrow C_\ell b_\ell^2+N_\ell \, ,
\eeq\label{clnoise}
where the beam function $b_\ell$ encodes the resolution limit of the instrument and $N_{\ell}$ is the noise power spectrum. The noise power spectrum for Planck was obtained from the publicly available 
SMICA \cite{PlanckXII:2013} noise map.
%
%
In addition to beam and noise effects, corrections to the power spectrum must also be made to account for the masking of the Galactic plane and point sources, among others, with the mask
$W(\hat{n})$. In Ref.~\cite{Hivon:2002} it has been shown that the measured power spectrum based on masked data can be corrected to obtained the unmaksed power spectrum as
\beq
\tilde{C}_\ell = \displaystyle\sum_{\ell'} M_{\ell \ell'} \mathcal{C}_{\ell'}
\eeq
where $M_{\ell \ell'}$ is defined by
\beq
M_{\ell \ell'} = \frac{2 \ell' + 1}{4\pi} \displaystyle\sum_{\ell \ell'} (2 \ell'' + 1) W_{\ell'}  \begin{pmatrix} \ell & \ell' & \ell'' \\ 0 & 0 & 0 \end{pmatrix}^2
\eeq\label{Ml}
when $W_{\ell}$ is the power spectrum of the mask $W(\hat{n})$. 
%
%


In order to obtain $\mathcal{K}_{\ell,conn}^{(2,2)}$, we created Gaussian simulations using the publicly available \texttt{healpix} software \cite{Gorski:2005} and applying Eq.~\ref{eq:g22} where $a_{\ell m}$ are obtained from Gaussian realizations of the Planck map. We then subtracted these simulations from the full trispectrum estimator $\mathcal{K}_{\ell,data}^{(2,2)}$ (Eq.~\ref{eq:k22data}) to obtain only the connected term. The full estimator and Gaussian piece are shown in Figure~\ref{fig:k22vg22}. After calculating $\mathcal{K}_{\ell,ana}^{(2,2)}$ (Eq.~\ref{eq:k22conn} with $C_{\ell}^{\alpha\alpha}$ set to 1), we estimated the power spectrum $C_{\ell}^{\alpha\alpha}$ by taking the ratio of the connected piece of the estimator from CMB data to the analytical connected piece. We obtain hence a $C_{\ell,sim}^{\alpha\alpha}$ for each of the gaussian simulations. The final $C_{\ell}^{\alpha\alpha}$ an its error bars are obtained averaging over all the $C_{\ell,sim}^{\alpha\alpha}$. Figure~\ref{fig:claa} shows the  angular power spectrum for spatial variations of $\alpha$, $C_{\ell}^{\alpha\alpha}$.  

%
%

%
%

As it can be seen in Figure~3 the measured  $C_{\ell}^{\alpha\alpha}$ is consistent with zero, showing no evidence for spatial variations of $\alpha$ when projected at the last scattering surface
at a redshift of 10$^3$. The most significant fluctuations are observed for the very low multipoles ($\ell<5$). However the value of $C_{\ell}^{\alpha\alpha}$ is always consistent with zero at the 
$2\sigma$ confidence level. We repeated the analysis described above removing the noise map from the original CMB map in order to show possible biasing effects due to the noise. The results are shown in Figure \ref{fig:claa} and we find that the noise bias is not affecting substantially the analysis.  Assuming $C_{\ell}^{\alpha\alpha}$  is a constant independent of $\ell$ the
 reduced $\chi^2$ fit of these two sets of data are 1.55 (SMICA) and 0.508 (SMICA, noise removed).

From the measured $C_{\ell}^{\alpha\alpha}$, we obtain the variance on $\alpha$ as
\beq
\sigma^2 = \frac{1}{4\pi} \displaystyle\sum_{\ell} (2\ell+1) C_{\ell}^{\alpha\alpha}
\eeq\label{eq:varalpha}
yielding 1$\sigma$ values of $\delta \alpha/\alpha_0 =  (0.668 \pm 2.91) \times 10^{-2}$ and $(0.379  \pm 2.91) \times 10^{-2}$ for SMICA and SMICA with noise removed, respectively, over the range of $2 < \ell < 20$,
corresponding to tens of degree angular scales and above. We find no evidence for a spatial variation of $\alpha$ at the quadrupole ($ell =2$) level and above at $z=10^3$. 
At redshifts probed by the galactic absorption line systems, a claim for a dipole variation of $\alpha$ exist at the 4 to 5 $\sigma$ level with a fluctuation amplitude of
$0.97 \pm 0.22 \times 10^{-5}$. The CMB measurements currently are three orders of magnitude worse than the capability of absorption line systems, but, has the ability to probe fluctuations 
over a wide range of angular scales. The CMB measurement we report here, unfortunately, is not sensitive to a dipole variation in $\alpha$ and thus we cannot test the existing result
from absorption line studies. Moreover, CMB probes the fluctuations at the last scattering surface at $z \sim 10^3$ significantly higher in redshift than absorption line systems. It could be that
$\alpha$ is evolving and the measurements at two epochs cannot be easily combined without an underlying model.


%
%


\begin{thebibliography}{199}


\bibitem{CODATA}
P.~J. Mohr,  B.~N. Taylor,~ and B.~N. Newell, Rev. Mod. Phys. 84, 1527  (2012) 
\bibitem{Webb98}
J. K. Webb, V. V. Flambaum, C. W. Churchill, et al., Phys. Rev. Lett 82, 884 (1999)
\bibitem{Webb01}
J. K. Webb, M. T. Murphy, V. V. Flambaum, et al.,Phys. Rev. Lett 87, 091301 (2001) 
\bibitem{Murphy:2003}
  M.~T.~Murphy, J.~K.~Webb, and V.~V.~Flambaum, Mon. Not. R. Astron. Soc. 345, 609 (2003)
\bibitem{Srianand:2004}
  R.~Srianand, H.~Chand, P.~Petitjean, and B.~Aracil, Phys. Rev. Lett 92, 121302 (2004)
\bibitem{Nakashima:2008}
M. Nakashima, R. Nagata and J. Yokoyama, Prog. Theor. Phys. 120, 1207 (2008) 
\bibitem{Martins:2004}
  C.~Martins, A.~Melchiorri, G.~Rocha, and R.~Trotta, Phys. Lett. B 585, 29, (2004)
 \bibitem{Menegoni:2012tq}
E.~Menegoni, M.~Archidiacono, E.~Calabrese, et al., Phys.\ Rev.\ D 85, 107301, (2012)
 \bibitem{Menegoni:2009rg}
 E.~Menegoni, S.~Galli, J.~G.~Bartlett, et al., Phys.\ Rev.\ D 80, 087302 (2009)
\bibitem{Rocha:2003}
G. Rocha, R. Trotta, C. J. A. Martins, et al., Mon. Not. Roy. Astron. Soc. 352, 20 (2004) 
\bibitem{Ichikawa:2006}
K. Ichikawa, T. Kanzaki and M. Kawasaki, Phys. Rev. D 74, 023515  (2006) 
\bibitem{Stefanescu:2007}
P. Stefanescu, New Astron. 12, 635 (2007)

\bibitem{Uzan:2003}
 J.~P.~Uzan,  Rev. Mod. Phys. 75, 403 (2003)
\bibitem{Bekenstein:2002}
  J.~D.~Bekenstein, Phys. Rev. D 66, 123514 (2002)
\bibitem{Youm:2002}
  D.~Youm, Phys. Lett. A 17, 175 (2002)
\bibitem{Sandvik:2002}
H.~B.~Sandvik, J.~D.~Barrow, and J.~Magueijo, Phys. Rev. Lett. 88, 031302 (2002)
 \bibitem{Murphy:2004}
M.~T.~Murphy, V.~V.~Flambaum, J.~K.~Webb, et al., Lect. Notes Phys. 648, 131 (2004)
\bibitem{King:2012} 
J.~A.~King, J.~K.~Webb, M.~T.~Murphy, et al., Mon. Not. R. Astron. Soc.  422, 4 (2012)
\bibitem{Hu:2001} 
W.~Hu, Angular trispectrum of the cosmic microwave background. 2001, Phys. Rev., D64, 083005, arXiv:astro-ph/0105117

\bibitem{Planck13} 
Planck Collaboration, arXiv:1303.5062 [astro-ph.CO].
\bibitem{Sigurdson:2009jr}
K.~Sigurdson, A.~Kurylov, and M.~Kamionkowski, Phys.\ Rev.\ D 68, 103509 (2003)
\bibitem{camb}  
A. Lewis, A. Challinor and A. Lasenby, Astrophys. J. 538, 473  (2000) 
\bibitem{Smidt:2011}
J. ~Smidt, A. Cooray, A. Amblard, et al.,   Astrophys.\ J.\  728, L1 (2011)
\bibitem{PlanckXII:2013}
Planck Collaboration (Ade, P.A.R. et al.) arXiv:1303.5072 [astro-ph.CO]
\bibitem{Hivon:2002}
E. Hivon, K.M. Gorski, C.B. Netterfield,  Astrophys.\ J.\ 5667, 2 (2002)
\bibitem{Gorski:2005}
K.M.~Gorski, E. Hivon, A.J.~Banday, et al., 2005, Astrophys.\ J. \ 622, 759 (2005)



\end{thebibliography}
\end{document}